\documentclass[aps,prx,twocolumn,superscriptaddress,floatfix,longbibliography,amsmath,amssymb,reprint,nobalancelastpage]{revtex4-2}

\usepackage{graphicx}
\usepackage{dcolumn}
\usepackage{bm}
\usepackage{hyperref}
\usepackage[dvipsnames]{xcolor}
\usepackage{upgreek}
\usepackage{comment}

\newcommand{\sublbl}[1]{(#1)}

\newcommand{\nc}{n_\text{c}}                
\renewcommand{\oc}{\omega_0}                
\newcommand{\ocInst}{\omega_c}              
\newcommand{\oL}{\omega_\text{L}}           
\newcommand{\nomDet}{\Delta_0}              
\newcommand{\Pin}{P_\text{in}}              
\newcommand{\Olin}{\Omega_{\text{m}}}                
\newcommand{\OlinN}[1]{\Omega_{\text{m},#1}}
\newcommand{\nmax}{n_\text{max}}
\newcommand{\xzpf}{x_\text{zpf}}
\newcommand{\dOspring}{\delta\Omega}          
\newcommand{\od}{\omega_\text{d}}           
\newcommand{\Oshifted}{\Omega}
\newcommand{\dONLchi}{\delta\Omega_\text{NL}}
\newcommand{\kcross}{k_{\leftrightarrow}}
\newcommand{\Jeff}{J_\text{NL}}
\newcommand{\rmi}{\mathrm{i}}

\renewcommand{\Re}{\operatorname{Re}}

\begin{document}
\title{Strong nanomechanical Duffing nonlinearity and interactions induced through cavity optomechanics}

\author{Jesse J. Slim}
\affiliation{Center for Nanophotonics, AMOLF, Science Park 104, 1098 XG Amsterdam, The Netherlands}
\affiliation{Australian Research Council Centre of Excellence in Quantum Biotechnology (QUBIC),
School of Mathematics and Physics, University of Queensland, St Lucia, Queensland 4072, Australia}
\author{Ewold Verhagen}
\affiliation{Center for Nanophotonics, AMOLF, Science Park 104, 1098 XG Amsterdam, The Netherlands}
 \email{verhagen@amolf.nl}

\date{\today}

\begin{abstract}
Nonlinearity is a key resource in both classical and quantum signal processing. 
Nonlinear nanomechanical elements have found applications ranging from sensing to computing, while networks of nonlinear resonators, as well as nonlinearly coupled networks of linear resonators, constitute promising platforms for simulating complex dynamics. 
Here, we experimentally demonstrate an approach to realizing  strong mechanical nonlinearity in nanomechanical resonators, fully controlled through optical laser drives. The mechanism exploits the nonlinearity of the radiation-pressure interaction in a cavity optomechanical system, which gives rise to a nonlinear optical spring effect.  
The resulting Duffing nonlinearity is conveniently tunable in strength via pump laser power, while its sign is controlled by laser detuning. 
Moreover, we demonstrate that the nonlinear optical spring mediates effective interactions between mechanical modes coupled to a common cavity, inducing tunable nonlinear interactions between them that impact spectral response and dynamics. 
These results establish cavity optomechanics as a versatile and in-situ reconfigurable platform for engineering nonlinear dynamics in resonators and networks.
\end{abstract}

\maketitle

Nonlinearity is a pervasive and often decisive feature of nanomechanical resonators~\cite{Lifshitz2009nonlinear}. Duffing-type dynamics give rise to phenomena such as frequency shifts, bistability, hysteresis, bifurcations, limit cycles and chaos~\cite{Guckenheimer1983nonlinear,Jordan2007nonlinear,Holmes1976bifurcations,Kovacic2011duffing}. It is also widely explored for applications in sensing, signal processing, and mechanical computing approaches~\cite{Mahboob2011interconnectfree,Mahboob2016electromechanical,Romero2024acoustically}. More broadly, exciting opportunities arise from controlled nonlinearity in both single resonators and networks or arrays: it can manipulate noise and fluctuations to yield strong squeezing~\cite{Huber2020spectral} and thermal engines~\cite{Serra-Garcia2016mechanical}, produce solitons and frequency combs~\cite{Kippenberg2018dissipative,Ochs2022frequency}, and enable novel computing paradigms including neuromorphic approaches and reservoir computing~\cite{Sun2021novel,Tanaka2019recent,Markovic2020physics}. In the quantum domain, nonlinearity enables quantum information processing and leads to intriguing correlated phases of matter~\cite{Hartmann2006strongly,Ludwig2013quantum}. But also in classical arrays with nonlinear interactions fascinating phases of matter and collective dynamics have been predicted and observed, including nonlinear and dynamical topological phases~\cite{Leykam2016edge,Meier2016observation,Mukherjee2020observation,Mukherjee2021observation,Lin2021dynamic,Kirsch2021nonlinear} and quantized nonlinear Thouless pumping~\cite{Jurgensen2021quantized,Jurgensen2023quantized}. Specifically, networks with nonlinear interactions have attracted interest, for example for enabling topological solitons that could occur at self-induced travelling domain walls \cite{Hadad2016selfinduced,Hadad2017solitons,Hadad2018selfinduced}.

Nanomechanics provides an excellent testbed for such dynamics, as well as a promising technological application platform. However, even if nanomechanical nonlinearities of geometrical origin are ubiquitous, they are generally weak, requiring significant amplitudes on the order of the dimensions of a nanostructure (e.g. the thickness of a flexural beam). Moreover, as they are rooted in 
material properties and device geometry, they exhibit very limited \emph{in-situ} tunability.

Here, we experimentally demonstrate a nanomechanical resonator with an optically tuneble, strong Duffing nonlinearity. Rather than from device geometry or material properties, this nonlinearity originates from cavity optomechanical backaction~\cite{Aspelmeyer2014cavity}: it is linked to the intrinsic nonlinearity of the cavity optomechanical interaction and fully induced and controlled through laser light. 
Previously, light-controlled Duffing nonlinearity was observed in dissipatively coupled optomechanical systems~\cite{Li2012multichannel,Westwood-Bachman2022isolation} for amplitudes of $\sim$100~nm and also predicted in coupled waveguides~\cite{Ashour2021spontaneous}. 
We show that nanomechanical nonlinearity can generally emerge from a standard dispersively coupled driven cavity optomechanical resonator, and can be orders of magnitude larger than intrinsic mechanical nonlinearity. In particular, the nonlinear shape of the cavity response as a function of displacement leads to a nonlinear optical spring effect, which results in tunable nonlinearities of different orders. We test this principle in an on-chip optomechanical device with exceptionally strong optomechanical coupling~\cite{Leijssen2017nonlinear}, so that motion explores a large fraction of the cavity response. We observe mechanical bistability for picometer-level vibrations, due to a Duffing nonlinearity whose sign and magnitude is completely controlled through an optical drive. Moreover, we demonstrate a scheme for realizing a pure inter-mode (cross-Kerr) nonlinearity between coupled resonators. 

\begin{figure*}
    \centering
    \includegraphics{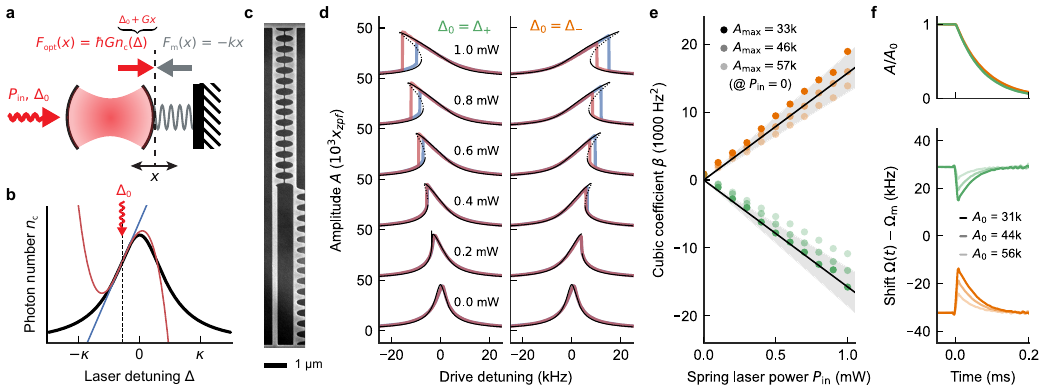}
    \caption{\textbf{Laser-controlled Duffing nonlinearity.} 
    \sublbl{a} Canonical optomechanical system: a cavity formed by two mirrors, where the movable right mirror experiences mechanical restoring force $F_\text{m}$ and radiation pressure $F_\text{opt}$. Mirror displacement $x$ linearly shifts the detuning $\Delta$ between cavity and pump laser. We assume the fast-cavity limit $\kappa \gg \Olin$, with $\kappa$ the cavity linewidth and $\Olin$ the intrinsic mechanical frequency.
    \sublbl{b} Cavity photon number $\nc$ depends nonlinearly on $\Delta$ (black), resulting in optical backaction $F_\text{opt}$ that is nonlinear in $x$. Tuning the nominal laser detuning $\Delta_0$ across the Lorentzian response controls this nonlinearity. At the indicated $\Delta_0 = -\kappa/(2\sqrt{3})$, a linear approximation (blue) for small displacements $Gx \ll \kappa$ gives the optical spring effect; for larger displacements $Gx \sim \kappa$, a cubic approximation (red) additionally yields Duffing nonlinearity. 
    \sublbl{c} Scanning electron micrograph of sliced nanobeam device. 
    Flexural modes of the suspended silicon beams couple dispersively to a broadband photonic crystal cavity. 
    \sublbl{d} Driven response of the mechanical resonance at $\Olin/2\pi = 17.6~$MHz for positive detuning $\Delta_+ = \kappa/(2\sqrt{3})$ (left) and negative detuning $\Delta_- = -\kappa/(2\sqrt{3})$ (right). Forward (blue) and backward (red) frequency sweeps are shown for increasing spring laser powers $\Pin$ (offset for clarity), with theory overlaid (black).
    \sublbl{e} Cubic coefficients $\beta$ for increasing $\Pin$ and drive amplitude $A_\text{max}$. Estimates from the experimental response (circles) match well with Eq.~\eqref{eq:beta_def} (black line) for both $\Delta_+$ (green, $\beta<0$) and $\Delta_-$ (orange, $\beta>0$).
    \sublbl{f} Ring-down at $\Pin=1.0$~mW for $\Delta_+$ (green, shift $> 0$) and $\Delta_-$ (orange, shift $<0$), for varying initial amplitude $A_0$. The nonlinearity does not affect the resonator's exponential amplitude decay (top), while it manifests in the evolving instantaneous frequency $\Omega(t)$ once the mechanical drive is switched off for $t>0$. 
    }
    \label{fig:1}
\end{figure*}

We consider a mechanical resonator with intrinsic resonance frequency $\Olin$ and mass $m$, whose displacement $x$ linearly shifts the resonance frequency $\ocInst = \oc - G x$ of an optical cavity 
with optomechanical coupling strength $G$, nominal frequency $\oc$ and linewidth $\kappa \gg \Olin$ (Fig.~\ref{fig:1}a). 
The latter condition---known as the fast-cavity or unresolved-sideband limit---implies that the cavity photon number $\nc$ responds instantaneously on mechanical timescales, such that photon dynamics can be adiabatically eliminated.
A pump 
laser with intensity $\Pin$ and frequency $\oL$, nominally detuned from the cavity by $\nomDet = \oL - \oc$, drives it to a mean photon number $\nc$. Under the nonlinear optomechanical interaction, 
the resonator displacement evolves as
\begin{align}
    m\ddot{x} = -k x - m \gamma \dot{x} + F_{\text{opt}}, \label{eq:mech_eom_1}
\end{align}
where $k = \Olin^2 / m$ is the mechanical spring constant, $\gamma$ its energy damping rate and $F_{\text{opt}} = \hbar G \nc$ the radiation pressure force exerted by the intracavity field. 

Importantly, $F_{\text{opt}}$ depends on the displacement $x$ as it shifts the instantaneous spring laser detuning $\Delta = \nomDet + G x$, which in turn modulates the intracavity photon number $\nc = \nmax h(2\Delta/\kappa)$
through the Lorentzian cavity response
\begin{align}
    h(u) = \frac{1}{1+u^2}. \label{eq:hu_lorentzian}
\end{align}
Here, we define the maximum photon number $\nmax = 4\Pin / (\hbar \oc \kappa)$ attained on resonance ($\Delta = 0$), and the relative detuning $u = 2\Delta / \kappa$. For small displacements $Gx \ll \kappa$, Eq.~\eqref{eq:hu_lorentzian} can be linearised around the average relative detuning $u_0 = 2\Delta_0 / \kappa$, yielding an optical backaction force
\begin{align}
    F_\text{opt} 
    &= F_\text{opt}^\text{DC} + \frac{2 \hbar G^2 \nmax h'(u_0)}{\kappa} x = F_\text{opt}^\text{DC} - k_\text{opt} x
\end{align}
that varies linearly with $x$. The static radiation pressure $F_\text{opt}^\text{DC} = \hbar G \nmax h(u_0)$ only displaces the resonator's equilibrium position $x_0$, which we account for by redefining $x \to x - x_0$.
Meanwhile, the optical spring constant $k_\text{opt}$ supplements the intrinsic mechanical spring $k$, shifting the mechanical resonance frequency by
\begin{align}
    \dOspring =
    \frac{k_\text{opt}}{2m\Olin}
    = -\frac{2 g_0^2 \nmax h'(u_0)}{\kappa}. \label{eq:dOmega_spring_shift}
\end{align}
This well-known shift is conveniently expressed using the vacuum optomechanical coupling rate $g_0 = G \xzpf$ and the resonator's zero-point quantum fluctuation amplitude $\xzpf = \sqrt{\hbar/(2m\Olin)}$, even though Eq.~\eqref{eq:dOmega_spring_shift} equally applies in the classical domain.

For larger amplitudes $Gx \sim \kappa$, the instantaneous detuning samples a significant fraction of the cavity response, so that the nonlinear shape of $h(u)$ needs to be taken into account. By moving the laser detuning $u_0$ across this Lorentzian, we tune the nonlinear terms in the optical force's Taylor series. To induce a cubic nonlinearity to leading order, we select $u_0 = u_\pm = \pm 1/\sqrt{3}$ (Fig.~\ref{fig:1}b) where $h''(u_0) = 0$. Consequently, $|h'(u_0)|$ is maximal, so that the linear spring shift $|\dOspring|$ is largest.
The equation of motion
\begin{align}
    \ddot{z} + \Oshifted^2 z + \gamma \dot{z} + \beta z^3 + \mathcal{O}(z^4) = f(t),\label{eq:duffing}
\end{align}
of the resonator's normalized displacement $z = x / \xzpf$ is then the \emph{Duffing equation}, featuring a cubic term with coefficient (see Supplementary Information)
\begin{align}
    \beta = -\frac{8\Olin \nmax h'''(u_\pm)g_0^4}{3\kappa^3} = - 6 \Olin \dOspring \frac{g_0^2}{\kappa^2}. \label{eq:beta_def}
\end{align}
In Eq.~(\ref{eq:duffing}), $\Oshifted = \Olin + \dOspring$ is the spring-shifted linear frequency of the mechanical resonator and $f(t)$ represents a driving force. We neglect the higher-order terms $z^n$ for $n>3$ in the optical force as their coefficients scale with $(g_0/\kappa)^{n} \ll 1 $.

Crucially, this optomechanically induced Duffing coefficient $\beta$ is fully tunable in both magnitude and sign. By setting $u_0 = -1/\sqrt{3}$, the linear spring shift $\dOspring < 0$ is negative, so that the corresponding $\beta > 0$ represents a hardening nonlinearity, while for $u_0 = 1/\sqrt{3}$ a softening nonlinearity $\beta < 0$ is obtained. Moreover, the magnitude of $\beta$ scales linearly with $\dOspring$ and is, therefore, directly tuned by the pump laser power $\Pin$.

In principle, this optically tuneable mechanical nonlinearity occurs in any fast-cavity optomechanical system. 
Whether it leads to observable effects depends on the vibrational amplitude. The critical phonon number $n_\text{NL}$ above which a coherently driven resonator's response becomes bistable~\cite{Schmid2023fundamentals} occurs when the nonlinear mechanical frequency shift exceeds the linewidth $\gamma$. It can be expressed as
\begin{align}
    n_\text{NL} = \frac{8}{3\sqrt{3}} \frac{\Olin \gamma}{|\beta|} = \frac{4}{9\sqrt{3}}\frac{\gamma}{|\dOspring|} \frac{\kappa^2}{g_0^2}.
\end{align}

Owing to its large ratio $g_0/\kappa$ and fast-cavity operation, a well-suited optomechanical system to observe nonlinearity at low amplitude is the sliced nanobeam photonic crystal cavity~\cite{Leijssen2015strong}, in which the effect of the nonlinear cavity response on the optical read-out of mechanical motion has been observed before~\cite{Leijssen2017nonlinear} as well as strong spring shifts~\cite{Mathew2020synthetic}.
We employ this sliced nanobeam platform to demonstrate a controlled Duffing nonlinearity. 
Our on-chip device features a suspended silicon beam with a mechanical resonance at $\Olin/2\pi = 17.6~$MHz (effective mass $m = 1$~pg) coupled to a telecom optical mode (linewidth $\kappa/2\pi=313$~GHz) at $g_0/2\pi = 3.5 \pm 0.3~$MHz. For this high-quality resonator ($\gamma/2\pi = 4.1$~kHz), we readily achieve spring shifts $|\dOspring| > 30$~kHz when illuminating the cavity with a focused `spring' laser of up to $\sim$1~ mW incident from free space. Here we remark that the cavity coupling efficiency is approximately 3\%, so the the coupled laser power is maximally a few tens of $\upmu$W. 

In Fig.~\ref{fig:1}d, we show the resonator's response to coherent driving $f(t) = f_0 \cos(\od t)$ at frequency $\od$, for a range of incident spring laser powers $\Pin$ and detunings $u_0$. The drive force $f(t)$, with fixed amplitude $f_0$, is applied optically by modulating an additional weak `force' laser resonant with the cavity, while a far-detuned `detect' laser reads out the resulting motion (see detailed methods description in Supplementary Information).
When the spring laser is off, we observe a Lorentzian response that is approximately symmetric in frequency, indicating essentially linear resonator dynamics for this drive force strength, apart from a small residual shift arising from the detection laser and intrinsic nonlinearity (see below).
Turning on the spring laser at detuning $\Delta_0=u_0\kappa/2 = \kappa/(2\sqrt{3})$, where $\dOspring > 0$, induces an observed softening nonlinearity that increases in strength with $\Pin$. The driving force is kept the same, so the optomechanically induced nonlinearity clearly exceeds the intrinsic mechanical nonlinearity of the beam. 
Notably, at higher $\Pin \ge 0.6$~mW, the resonator exhibits the bistable `shark fin' response typical for a Duffing resonator, and associated hysteresis depending on the sweep direction of $\od$. Conversely, for red-detuned $\Delta_0=u_0\kappa/2 = -\kappa/(2\sqrt{3})$, where $\dOspring < 0$, we find a hardening nonlinearity of similar strength but opposite sign. In that case, the characteristics of the response appear reflected in frequency. At both detunings, the experimental frequency response agrees well with Duffing theory (black curves in Fig.~\ref{fig:1}d, see Supplementary Information) for the coefficient $\beta$ predicted by Eq.~\eqref{eq:beta_def}.

Evidently, the nonlinear resonance shift $\sim \beta A^2/\Olin$ opposes the linear spring shift $\dOspring$.
This can be understood intuitively: at $u_0 = \pm 1/\sqrt{3}$ the derivative $|h'(u_0)|$ is maximal, so that any detuning excursions will reduce the average derivative of the cavity response and, thus, the effective spring shift. Ultimately, for extremely large amplitudes $GA \gg \kappa$, the cavity will be far off-resonant for most of the mechanical cycle, so that the resonator reverts back to its intrinsic resonance frequency (in the absence of other nonlinearities).

The cubic coefficient $\beta$ can be extracted from the experimental data, by inverting the relation $A_\text{max} = \sqrt{8(\omega_\text{d,max} - \Oshifted)\Oshifted/(3\beta)}$ between 
the optimal drive frequency $\omega_\text{d,max}$ and amplitude $A_\text{max}$ of the maximum response~\cite{Schmid2023fundamentals}.
As shown in Fig.~\ref{fig:1}e, experimentally obtained estimates of $\beta$ match well with Eq.~\eqref{eq:beta_def}, showcasing the full tunability of our scheme. 
At zero pump power, a small residual nonlinearity $\beta < 900$~Hz$^2$ is estimated from the observed shift in the response peak. This is related to the intrinsic nonlinearity as well as the optomechanical nonlinearity of the detuned detection laser, which remains on even for zero spring laser power. We estimate the intrinsic geometrical nonlinearity contribution at $\beta_\text{intr} \approx 300$~Hz$^2$ from finite element modelling (Supplementary Information) and the contribution from the detection laser at $\beta_\text{det} \approx -30$~Hz$^2$. 

The largest observed optomechanical nonlinear coefficient $\beta \approx 16 000$~Hz$^2$ is thus $50$ times larger than the strength of the intrinsic nonlinearity. 
Appreciable Duffing nonlinearity emerges at critical amplitude $A_\text{NL}=\sqrt{n_\text{NA}}\xzpf$ as low as $A_\text{NL} = 17\,000\,\xzpf$, only $20$ times larger than the thermal amplitude of our room-temperature resonator. Moreover, with $\xzpf = 22$~fm (estimated from simulations), $A_\text{NL}=0.36$~nm represents a lateral displacement of less than a nanometer, much smaller than the critical amplitudes observed through different schemes~\cite{Li2012multichannel} or typical intrinsic mechanical nonlinearities for flexural nanomechanical resonators of similar dimensions.

We probe the time-domain dynamics of our nonlinear resonator in a ring-down experiment (Fig.~\ref{fig:1}f). The resonator is first excited to a high amplitude $A_0$ by a spring laser modulation, which is subsequently switched off at $t=0$. 
Initially, for $t<0$, the instantaneous frequency $\Omega(t)$ of the resonator's motion (see Supplementary Information), shown in Fig.~\ref{fig:1}f, bottom, is locked to the drive. After the modulation is switched off, the resonator is free to evolve at its natural frequency, which for a Duffing resonator depends on its amplitude~\cite{Jordan2007nonlinear,Kovacic2011duffing}. Consequently, at $t=0$, the estimated instantaneous frequency $\Omega(t)$ jumps to a nonlinearly shifted value. As the resonator's amplitude rings down (top), this nonlinear shift gradually reduces, until the resonator oscillates at its linear (spring-shifted) frequency $\Oshifted$.

\begin{figure}[tb]
    \centering
    \includegraphics{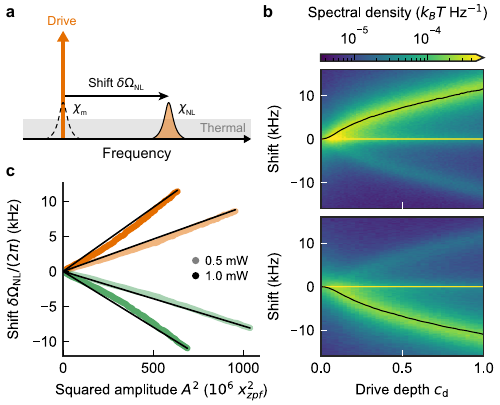}
    \caption{\textbf{Nonlinear shift in susceptibility to thermal fluctuations.}
    \sublbl{a} A strong mechanical drive induces large amplitude oscillations, shifting the resonator's susceptibility $\chi_\text{NL}$ to additional (thermal) forces.
    \sublbl{b} 
    Thermomechanical spectra of the resonator, coherently driven by modulations of a weak, resonant `force' laser at increasing depth $c_\text{d}$.
    The strong ($\Pin = 1.0$~mW) spring laser is detuned by $\Delta_- = -\kappa/(2\sqrt{3})$ (top) or $\Delta_+ = \kappa/(2\sqrt{3})$ (bottom). Bright bands indicate thermal fluctations on top of the driven amplitude, while fainter bands result from nonlinear transduction. 
    \sublbl{c} Fitted center frequencies of the Lorentzian thermal contribution plotted against the square of the driven amplitude, for $\Delta_-$ (orange) and $\Delta_+$ (green) at two powers $\Pin$.
    Expected susceptibility shifts from Eq.~\eqref{eq:chiNL_shift} (black) are overlaid in \sublbl{b} and \sublbl{c}.
    }
    \label{fig:2}
\end{figure}

We continue by exploring the resonator's thermal dynamics in the presence of nonlinearity. Figure~\ref{fig:2} shows the measured thermomechanical spectrum of the resonator while it is simultaneously driven by a force laser modulation at increasing depth $c_\text{d}$. The coherent high-amplitude oscillations appear as a narrow peak with increasing amplitude, saturating the color scale. Notably, the broad spectrum of thermal vibrations now shifts in frequency, with the direction of the shift depending on the spring laser detuning: for $u_0 = -1/\sqrt{3}$ it moves up in frequency as $c_\text{d}$ increases, while for $u_0 = 1/\sqrt{3}$ it moves down. 
We note that additionally a weak, frequency-reflected ‘ghost’ band appears that moves in the other direction. We attribute this to a combination of nonlinear transduction~\cite{Leijssen2017nonlinear} that mixes the thermal fluctuations with the strong coherent oscillations, or a small magnitude of nonlinearly driven motion at such frequencies, as observed before in nonlinear electromechanical~\cite{Huber2020spectral} and microwave~\cite{FaniSani2021level} systems.

Figure~\ref{fig:2} demonstrates that in the presence of optomechanical nonlinearity, coherent oscillation alters the resonator's susceptibility $\chi_\text{NL}$ to small thermal forces. Using harmonic balance~\cite{Krack2019harmonic} (Supplementary Information), we find that $\chi_\text{NL}$ remains Lorentzian like its linear counterpart $\chi_\text{m}$ (Fig.~\ref{fig:2}a), but with a center frequency shifted by 
\begin{align}
    \dONLchi = \frac{3}{4} \frac{\beta A^2}{\Olin}= -\frac{9}{2} \dOspring \frac{g_0^2}{\kappa^2}A^2, \label{eq:chiNL_shift}
\end{align}
where $A$ is the normalized amplitude of the strong coherent oscillations. The thermal peak frequencies fitted from Fig.~\ref{fig:2}b, plotted in Fig.~\ref{fig:2} against the measured coherent amplitude $A$, agree well with the predicted values from Eq.~\eqref{eq:chiNL_shift}.

\begin{figure*}
    \centering
    \includegraphics{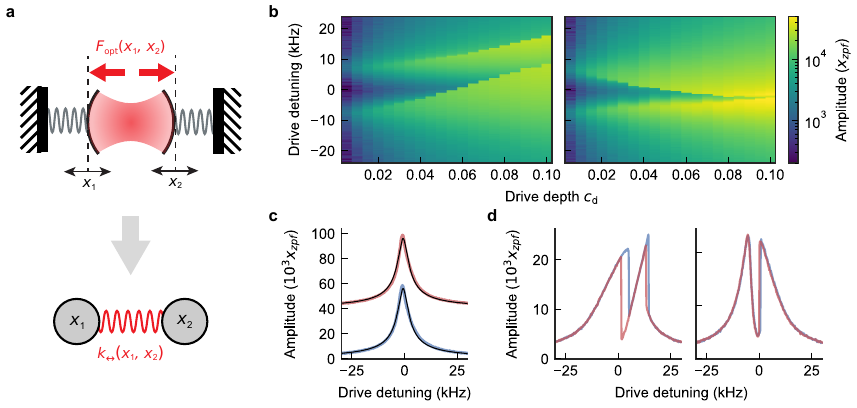}
    \caption{\textbf{Optically-mediated nonlinear coupling between mechanical resonators.}
    \sublbl{a} Multi-mode optomechanical system: both mirrors are free to move, so that photon number $n_c$ depends nonlinearly on both displacements $x_1, x_2$ via the Lorentzian cavity response. 
    The resulting radiation pressure $F_\text{opt}$ on each mirror induces a nonlinear, amplitude-dependent optical spring $\kcross(x_1, x_2)$ coupling the two mechanical elements.
    \sublbl{b, left} Forward-sweep frequency response of resonator $2$ ($\OlinN2/2\pi = 12.8$~MHz) for increasing drive depth $c_\text{d}$, while coupled to resonator $1$ ($\OlinN1/2\pi = 17.6$~MHz) via a cross-resonator optical spring induced by modulating a single spring laser (detuning $\Delta_- = -\kappa/(2\sqrt{3})$) at their difference frequency $\Delta\Omega_{12}$. 
    \sublbl{b, right} Similar, but with an additional static spring laser at detuning $\Delta_+ = \kappa/(2\sqrt{3})$ that cancels the intra-resonator nonlinearity while preserving the cross-resonator nonlinearity, resulting in reduced peak splitting with increasing amplitude.
    \sublbl{c} Response of uncoupled resonator $1$ with both spring lasers active. Cancellation of nonlinearity restores a Lorentzian lineshape and eliminates hysteresis between forward (blue) and backward (red) sweeps (offset for clarity).
    \sublbl{d} Response of coupled resonator $2$ at fixed drive depth, for single-laser (left, $c_\text{d} = 0.08$) and two-laser (right, $c_\text{d} = 0.045$) driving. Hysteresis between forward (blue) and backward (red) sweeps is observed for single-laser driving, yet largely absent for two-laser driving.
    }
    \label{fig:3}
\end{figure*}

The above results demonstrate how nonlinear optical backaction can control the nonlinear dynamics of just a single mechanical mode. Interestingly, rich multimode nonlinear dynamics can be induced for multiple resonators that are simultaneously coupled to a common cavity. In that case, the cross-resonator optical spring, produced by their mutual backaction, can generate effective nonlinear hopping interactions between the mechanical modes. This has the prospect of creating cross-Duffing interactions, analogous to cross-Kerr effect in optics and photonics. Such interactions are essential for a variety of fascinating phenomena, including nonlinear energy localization~\cite{Sato2006nonlinear}, frequency pulling in synchronization~\cite{Matheny2014phase}, correlated phases of matter~\cite{Jin2013photon} and nonlinear dynamical topological solitons~\cite{Hadad2016selfinduced,Hadad2017solitons}. 

Here we analyse such coupled nonlinear dynamics. Two cavity-coupled resonators with displacements $x_1,x_2$ (Fig.~\ref{fig:3}a) modulate the instantaneous cavity detuning $\Delta = \Delta_0 + G_1 x_1 + G_2 x_2$ through their optomechanical coupling rates $G_j$. Consequently, the optical force $F_{\text{opt},j}(x_1, x_2)$ acting on each resonator $j$ now depends on \emph{both} their displacements. For small amplitudes, linearizing $F_{\text{opt},j}(x_1, x_2)$ yields a cross-resonator optical spring that dynamically couples $x_1 \leftrightarrow x_2$ by a linear spring constant $k_\leftrightarrow$ (see Supplementary Information).
However, for large displacements $G_1 x_1, G_2 x_2 \sim \kappa$, the nonlinear cavity response must again be taken into account, resulting in a displacement-dependent cross-resonator spring $\kcross(x_1, x_2)$.

The sliced photonic crystal nanobeam hosts several flexural mechanical overtones (Supplementary Information) that couple to the cavity, allowing us to study such nonlinear coupling. We focus on a pair of modes that includes the aforementioned $\OlinN1/2\pi = 17.6$~MHz mode and a lower-frequency mode at $\OlinN2/2\pi = 12.8$~MHz with comparable damping rate $\gamma_2/2\pi = 3.7$~kHz and vacuum optomechanical coupling rate $g_{0,2}/2\pi = 3.8 \pm 0.1$~MHz. With the spring laser, both modes experience spring shifts up to $|\dOspring_j| \sim 30~$kHz so that they resonate at $\Oshifted_j = \OlinN{j} + \dOspring_j$. 
Due to the large frequency separation $\Delta\Omega_{12} = \Oshifted_1 - \Oshifted_2 \gg \gamma_j$, the static cross-resonator optical spring $\kcross$ has virtually no effect on their dynamics. That frequency difference can, however, be overcome by modulating the intensity $\Pin$ of the spring laser at $\Delta\Omega_{12}$. The resulting time-modulated spring $\kcross(t)$ stimulates frequency conversion between the modes, by producing sidebands in the optical force that are mutually resonant. For small amplitudes, this leads to an effective linear hopping interaction that resonantly couples both modes at a rate $J = c_\text{h} \sqrt{\dOspring_1 \dOspring_2} / 2$ (Supplementary Information), where $c_\text{h}$ is the depth of the spring laser modulation~\cite{Mathew2020synthetic}. 

In Fig.~\ref{fig:3}b (left panel), we probe the coherent response of the lower-frequency resonator $2$ in the presence of a coupling modulation with depth $c_\text{h} = 0.5$, imprinted on the red-detuned ($u_0 = -1/\sqrt{3}$) spring laser. Here, resonator $2$ is driven around its resonance frequency $\Oshifted_2$ by an additional weak modulation of the spring laser at depth $c_\text{d}$. For small $c_\text{d} < 0.02$, two Lorentzian peaks are observed split by frequency $2J$, signifying hybridisation of the two mechanical resonators by the effective optical coupling. For stronger drives, the larger amplitude induces nonlinearity both in the individual optical springs $k_{\text{opt},j}$, as well as the cross-resonator spring $\kcross$. The former increases the effective mechanical resonance frequencies at higher amplitudes, illustrated in Fig.~\ref{fig:3}b, top, by the overall upshift in peak frequencies, while the latter reduces their effective splitting, which is proportional to the (amplitude-dependent) coupling rate.

Conveniently, the cross-Duffing nonlinearity can be isolated by simultaneously applying \emph{two} spring lasers at opposite detunings $u_\pm = \pm 1/\sqrt{3}$ to cancel the cubic nonlinearity of the individual resonators.
As shown in Fig.~\ref{fig:3}c, a symmetric mechanical response is then restored that exhibits no hysteresis, even at high drive power. 
By again modulating only one of the spring lasers at $\Delta\Omega_{12}$, the nonlinearity in the cross-resonator spring $\kcross$ is, nevertheless, induced (Supplementary Information). As presented in Fig.~\ref{fig:3}b (right panel), the response of resonator 2 then remains centered around $\Omega_2$ with increasing $c_\text{d}$. For larger amplitudes, we do, however, see a strong reduction in the frequency splitting between the peaks, indicating a reduction in $\kcross$ by the nonlinearity. 
This is also reflected in the example drive frequency response curves shown in Fig.~\ref{fig:3}d: For a single modulated laser, these resemble two shark-fin-like curves for the two hybridized normal modes of the pair. With an isolated cross-Duffing nonlinearity, the response spectrum looks more symmetric, characterized by two peaks that tilt towards each other; shifting closer together as the amplitude increases.

\begin{figure}
    \centering
    \includegraphics{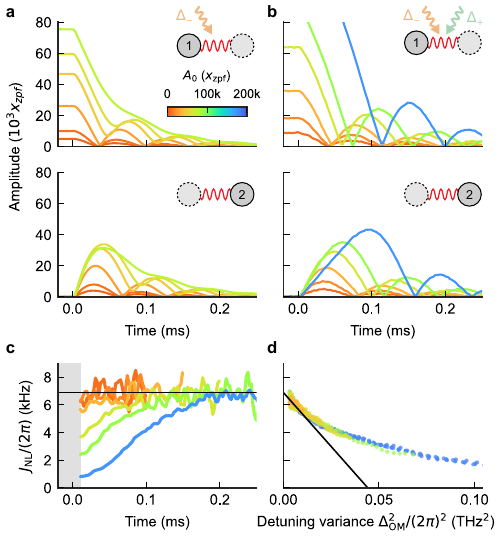}
    \caption{\textbf{Time-evolution of nonlinearly coupled resonators.}
    \sublbl{a} Ring-down of coupled resonators $1$ (top) and $2$ (bottom) under single-laser driving at detuning $\Delta_-=-\kappa/(2\sqrt{3})$. For $t<0$, resonator $1$ is coherently excited by modulating the spring laser at depth $c_\text{d}$. At $t=0$, the drive tone is switched off and a coupling tone at the difference frequency $\Delta\Omega_{12}$ is switched on, inducing Rabi oscillations whose period increases and visibility reduces at larger amplitudes.
    \sublbl{b} Similar, but with an additional static spring laser at detuning $\Delta_+ = \kappa/(2\sqrt{3})$. 
    Cancellation of intra-resonator nonlinearity keeps the coupling tone resonant to restore Rabi oscillation visibility, while the retained cross-resonator nonlinearity isolates the amplitude-dependent period lengthening.
    \sublbl{c} Instantaneous coupling rates $J_\text{NL}(t)$ extracted from the coupled ring-downs in \sublbl{b}. For small drive amplitudes $c_\text{d}$, estimated $J_\text{NL}(t)$ approach the linear coupling $J/2\pi \approx 7$~kHz (black) throughout. For larger drives, $J_\text{NL}$ is initially reduced, recovering toward $J$ as the amplitude dissipates. 
    \sublbl{d} Plotting the estimated $J_\text{NL}$ against the detuning variance induced by both resonators reveals a clear trend independent of initial drive amplitude, with good agreement with Eq.~\eqref{eq:J_NL} (black) for small to moderate mechanical amplitudes.
    }
    \label{fig:4}
\end{figure}

Nonlinear coupling also impacts the resonator pair's temporal dynamics. In the ringdown experiments shown in Fig.~\ref{fig:4}a for a single spring laser at $u_0=-1/\sqrt{3}$, resonator $1$ is resonantly driven until $t=0$, when the coupling tone at $\Delta\Omega_{12}$ is switched on. For small amplitudes, we observe Rabi oscillations in which energy is exchanged between the resonators at the Rabi frequency $J$ during ringdown. For larger driving amplitude $c_\text{d}$, the Rabi period increases --- indicating a nonlinear reduction in the effective coupling rate $\Jeff$ --- while visibility reduces. The latter results from unequal intra-resonator nonlinearities, which detunes their effective frequency difference from the coupling tone at $\Delta\Omega_{12}$ causing incomplete Rabi oscillations.

With two spring lasers at $u_\pm$ (Fig.~\ref{fig:4}b), a pure cross-resonator nonlinearity is isolated. The coupling tone then remains resonant, so that full fringe visibility is recovered at high amplitudes. The effective Rabi rate $\Jeff$ clearly increases as the resonator amplitudes ring down. Figure~\ref{fig:4}c quantifies this behavior by extracting the instantaneous value of $\Jeff$ from the phase and amplitude of both resonators (Supplementary Information). We see that $\Jeff$ starts out at a reduced effective coupling rate at $t=0$ and then gradually increases to $J/2\pi\approx 7$~kHz as the overall amplitude dissipates. Finally, Fig.~\ref{fig:4}d plots the extracted values of $\Jeff$ against the summed squared amplitude of both resonators, expressed as the variance in the optomechanical detuning shift 
$\Delta_\text{OM}^2 = g_{0,1}^2 a_1^2 + g_{0,2}^2 a_2^2$, where $a_j = A_j/x_{\text{zpf},j}$ is the normalized amplitude. We see a clear trend independent of the initial amplitude $A_0$, which, for small $\Delta_\text{OM}^2$ agrees well with the theoretically predicted
\begin{align}
    \Jeff = J \left[1 - \frac{9}{4} \left( \frac{g_{0,1}^2}{\kappa^2} a_1^2 + \frac{g_{0,2}^2}{\kappa^2} a_2^2 \right) \right] \label{eq:J_NL}
\end{align}
that we derive in the Supplementary Information. These results illustrate the capability of optomechanical backaction to precisely control nonlinear coupling rates in multimode mechanical systems.

In conclusion, we have experimentally demonstrated that cavity optomechanical backaction can serve as a powerful and highly tunable source of nanomechanical nonlinearity. In the fast-cavity regime, the nonlinear Lorentzian response of the optical cavity produces a nonlinear optical spring that gives rise to strong Duffing dynamics. Using a sliced photonic crystal nanobeam cavity, we used this mechanism to realize optically induced nonlinearities that exceed the intrinsic mechanical nonlinearity by more than an order of magnitude, yielding sub-nanometer level critical nonlinear amplitudes. We note that the mechanism is very general, and will occur in \emph{any} cavity optomechanical resonator in the fast-cavity regime with standard dispersive optomechanical coupling. Even the current sliced nanobeam platform could be improved to yield still much stronger nonlinearity, as $g_0$ values have been observed that exceed that in the current demonstration by a factor 10, and optical linewidths $\kappa$ that are at least one order of magnitude smaller~\cite{Leijssen2017nonlinear}. As the Duffing coefficient $\beta$ scales with $g_0^4/\kappa^3$, it is expected to be many orders of magnitude larger for optimized parameters. We do note that that increases higher-order nonlinearities by an even greater amount, resulting in an interesting trade-off depending on the specific target application.

A key feature of this nonlinearity is that it is fully controllable through the power and detuning of external laser drives. We showed that the sign and magnitude of both self-Duffing and cross-Duffing interactions can be individually tuned by using multiple lasers. Moreover, specific interactions can be addressed in the synthetic mode dimension through judicious temporal modulation. This points to the possibility of engineering arbitrary nonlinear Hamiltonians in networks of resonators through spectral control of drive fields at frequency scales of the order of optical linewidths and mechanical frequencies. Interactions can be dynamically reconfigured in situ fully optically, without requiring changes to device geometry or material composition. Indeed, while we have focused here on the third-order Duffing nonlinearity, suitable laser drives could generate and optimize different nonlinearity orders related to the spectral shape of the optical force.

Such control could be exploited in a variety of contexts. At the level of individual resonators, optically induced Duffing nonlinearities may enable low-power bifurcation devices, nonlinear sensing schemes, mechanical logic elements, and tunable susceptibilities for noise manipulation and squeezing. In multimode systems, engineered nonlinear couplings provide a promising platform for exploring collective nonlinear dynamics, synchronization, nonlinear transport, and driven-dissipative phases in coupled oscillator networks~\cite{Ludwig2013quantum,Leykam2016edge,Meier2016observation}. In particular, the realization of controllable cross-Duffing interactions could enable nonlinear topological phenomena and soliton formation in optomechanical lattices~\cite{Hadad2016selfinduced,Hadad2017solitons}. In the quantum regime, nonlinear backaction provides a powerful approach to the engineering of nonclassical quantum states and mechanical qubits~\cite{Clarke2025deterministic,Samanta2023nonlinear}.

More broadly, these results establish cavity optomechanics as a versatile framework for realizing reconfigurable nonlinear dynamics at the nanoscale. Combined with the scalability of nano-optomechanical architectures and the possibility of measurement and control down to the quantum regime, this provides a powerful setting for exploiting controlled nonlinearity for fundamental studies as well as technological applications of nonlinear resonators and networks.

\bibliography{om_duffing_export}


\clearpage
\pagebreak
\renewcommand{\theequation}{S\arabic{equation}}
\renewcommand{\thefigure}{S\arabic{figure}}
\setcounter{equation}{0}
\setcounter{table}{0}
\renewcommand{\figurename}{FIG.}
\setcounter{figure}{0}
\renewcommand{\theHfigure}{S.\thefigure}



\newpage
\onecolumngrid
\part*{\large\centering Supplementary Information}

\twocolumngrid

\section{Nonlinear optical backaction}
In this section, we present a detailed analysis of optical backaction in fast-cavity optomechanical systems~\cite{Aspelmeyer2014cavity}. We start from the equation of motion
\begin{align}
    m\ddot{x} = -k x - m \gamma \dot{x} + F_{\text{opt}}, \label{eq:SI:mech_eom}
\end{align}
for a mechanical resonator subject to radiation pressure force $F_{\text{opt}}$, given in the main text as Eq.~\eqref{eq:mech_eom_1}. 
Optical backaction arises as the optical force 
\begin{align}
    F_{\text{opt}} = \hbar G \nc = \hbar G \nmax h\left(\frac{2\Delta_0}{\kappa} + \frac{2 G x}{\kappa}\right) 
\end{align}
depends instantaneously on the mechanical displacement $x$, through the optomechanical coupling $G$ and the Lorentzian cavity response $h(u)$. This treatment is valid in the fast-cavity limit where the cavity linewidth $\kappa \gg \Olin$ is much larger than the mechanical frequency, so that the photon dynamics can be eliminated adiabatically. 

To analyse the nonlinear relation between $F_{\text{opt}}$ and $x$, we expand $h(u)$ into its Taylor series around $u_0 = 2\Delta_0/\kappa$ and obtain
\begin{align}
    &F_{\text{opt}} = \hbar G \nmax \Bigg[ h(u_0) + h'(u_0) \left(\frac{2 G x}{\kappa} \right) \label{eq:SI:Fopt_taylor} \\
    &\quad+ \frac{h''(u_0)}{2} \left(\frac{2 G x}{\kappa} \right)^2 + \frac{h'''(u_0)}{6} \left(\frac{2 G x}{\kappa} \right)^3 + \mathcal{O}(x^4) \Bigg]\nonumber
\end{align}
The constant term $F_\text{opt}^\text{DC} = \hbar G \nmax h(u_0)$ represents a static radiation pressure, which displaces the resonator's equilibrium position $x_0$. Conventionally, this is accounted for by redefining $x \to x-x_0$.

The linear term 
\begin{align}
    F_\text{opt}^{(1)} = \frac{2 \hbar \nmax h'(u_0) G^2}{\kappa} x = - k_\text{opt} x
\end{align}
gives rise to the conventional, linear optical spring effect, characterised by spring constant $k_\text{opt}$ that supplements the intrinsic mechanical spring constant $k = m\Olin^2$. It shifts the mechanical frequency by
\begin{align}
    \dOspring &= \Oshifted - \Olin = \sqrt{\frac{k + k_\text{opt}}{m}} - \sqrt{\frac{k}{m}} \\
    &= \Olin \left( \sqrt{1 + k_\text{opt}/k} - 1 \right) \\
    &\approx \frac{\Olin k_\text{opt}}{2k} = \frac{k_\text{opt}}{2m\Olin},
\end{align}
where the approximation is valid for small optical spring $k_\text{opt} \ll k$. Plugging in $k_\text{opt}$ yields
\begin{align}
    \dOspring &= -\frac{\hbar \nmax h'(u_0) G^2}{m\Olin \kappa} = -\frac{2 \nmax h'(u_0) g_0^2}{\kappa},
\end{align}
where we have substituted the vacuum optomechanical coupling rate $g_0 = G\xzpf$ and zero-point fluctuation amplitude $\xzpf = \sqrt{\hbar/(2m\Olin)}$. This relation for the linear optical spring shift is given in the main text as Eq.~\eqref{eq:dOmega_spring_shift}.

Next, we express the mechanical equation of motion using the normalized displacement $z = x / \xzpf$,
\begin{align}
    \ddot{z} = -\Olin^2 z - \gamma \dot{z} + f_{\text{opt}}(z), \label{eq:SI:mech_eom_z}
\end{align}
where $f_{\text{opt}}(z) = F_{\text{opt}}/(m\xzpf)$ is the normalized optical force. For reference, we give the linear term in the normalized force as
\begin{align}
    f_\text{opt}^{(1)} = \frac{4 \Olin \nmax h'(u_0) g_0^2}{\kappa} z = -2\Olin\dOspring \,z
\end{align}

The quadratic term in Eq.~\eqref{eq:SI:Fopt_taylor},
\begin{align}
    F_\text{opt}^{(2)} &= \frac{2 \hbar G^3 \nmax h''(u_0)}{\kappa^2} x^2,  
\end{align}
yields a contribution 
\begin{align}
    f_\text{opt}^{(2)} = \frac{4 \Olin \nmax h''(u_0) g_0^3}{\kappa^2} z^2 = -\alpha z^2
\end{align}
to the normalized optical force. However, precisely at the spring laser detunings $u_0 = u_\pm = \pm 1/\sqrt{3}$ that we use, where the optical spring shift $\dOspring$ is maximally positive or negative, respectively, the second derivative $h''(u_\pm)$ vanishes. Consequently, the quadratic nonlinear coefficient $\alpha$ induced by the spring laser is zero in all of our experiments.

In contrast, the cubic term in Eq.~\eqref{eq:SI:Fopt_taylor},
\begin{align}
    F_\text{opt}^{(3)} &= \frac{8\hbar \nmax h'''(u_0) G^4}{6\kappa^3} x^3,
\end{align}
does not vanish at $u_0 = u_\pm$. When normalized, the cubic optical force term reads
\begin{align}
    f_\text{opt}^{(3)} &= \frac{8 \Olin \nmax h'''(u_0) g_0^4  }{3\kappa^3} z^3 = -\beta z^3,
\end{align}
where $\beta$ is the cubic nonlinear coefficient.
This coefficient can be conveniently written using the linear spring shift $\dOspring$, as
\begin{align}
    \beta = \Olin \dOspring \frac{4 h'''(u_0)}{3 h'(u_0)} \frac{ g_0^2  }{\kappa^2}.
\end{align}
This expression is valid for any spring laser detuning $u_0$. At $u_0 = u_\pm$, we find $h'''(u_\pm) / h'(u_\pm) = -9/2$, so that we can write
\begin{align}
    \beta = -6\Olin\dOspring \frac{ g_0^2  }{\kappa^2} \label{eq:SI:beta_from_spring},
\end{align}
given in the main text as Eq.~\eqref{eq:beta_def}.

\section{Frequency response of a Duffing resonator}
In this section, we analyse the frequency response of a Duffing resonator to a strong `pump' force, optionally supplemented by a weak `probe' force. We do so using the method of harmonic balance~\cite{Krack2019harmonic}. This approach, commonly used to evaluate the dynamics of Duffing oscillators~\cite{Jordan2007nonlinear}, relies on expanding the displacement $z(t)$ as a Fourier series and then truncating that expansion to a specified order, while balancing the amplitudes of the remaining harmonics.

We start from the equation of motion
\begin{align}
    \ddot{z} + \gamma \dot{z} + \Oshifted^2 z + \beta z^3 = f(t) \label{eq:SI:freq_resp_eom1}
\end{align}
that describes the evolution of a general Duffing oscillator subject to a two-tone driving force
\begin{align}
    f(t) = f_1 \cos(\omega_1 t + \psi_1) + f_2 \cos(\omega_2 t + \psi_2)
\end{align}
that includes a strong `pump' tone with amplitude $f_1$ and frequency $\omega_1$ and a weak `probe' tone with amplitude $f_2$ and frequency $\omega_2$.
For simplicity, we assume that the frequencies $\omega_1, \omega_2$ are incommensurate, such that the phase offsets are not essential and we can set $\phi_1 = \phi_2 = 0$ without affecting the dynamics.

In response to this composite force, we seek an approximate solution to the resonator evolution of the form
\begin{align}
    z(t) = a_1 \cos(\omega_1 t + \phi_1) + a_2 \cos(\omega_2 t + \phi_2). \label{eq:SI:freq_resp_z_ansatz}
\end{align}
We insert Eq.~\eqref{eq:SI:freq_resp_z_ansatz} into the Duffing equation~\eqref{eq:SI:freq_resp_eom1} and balance the fundamental harmonics at $\omega_1$ and $\omega_2$. First, we solve for the pump amplitude $a_1$ while neglecting the probe amplitude ($a_2=0$). This results in the well-known equation
\begin{align}
    a_1^2 \left[\left( \omega_1^2 - \Omega^2 - \frac{3}{4}\beta a_1^2 \right)^2 + (\gamma\omega_1)^2 \right] = f_1^2 \label{eq:SI:duffing_response}
\end{align}
that describes the approximate frequency response of a Duffing oscillator to harmonic driving. This equation underlies the theory curves plotted in Fig.~\ref{fig:1}d. For driving $f_1 < f_1^c$ below the critical driving amplitude, Eq.~\eqref{eq:SI:duffing_response} admits a single solution for the amplitude $a_1$, while for strong driving $f_1 > f_1^c$ the amplitude becomes multi-valued and hysteresis emerges. 

Next, we solve for the probe amplitude $a_2$ in the presence of the pump. After balancing the fundamental harmonics, we find
\begin{align}
    a_2^2\left[\left( \omega_2^2 - \Omega^2 - \frac{3}{4}\beta(2a_1^2 + a_2^2)\right)^2 + (\gamma\omega_1)^2 \right] = f_2^2
\end{align}
Assuming that $a_1 \gg a_2$ in the term in brackets, we arrive at the approximation
\begin{align}
    a_2^2 = \frac{f_2^2}{(\mu^2 - \omega_2^2)^2 + \gamma^2 \omega_2^2 }
\end{align}
for the squared amplitude of motion at the probe frequency. This corresponds exactly to the susceptibility of a linear harmonic oscillator with natural frequency $\mu$. The shifted frequency $\mu$ satisfies the expression
\begin{align}
    \mu^2 = \Omega^2 + \frac{3}{2}\beta a_1^2.
\end{align}
After plugging in Eq.~\eqref{eq:SI:beta_from_spring} for the radiation-pressure-induced $\beta$, we arrive at the expression
\begin{align}
    \dONLchi = \mu - \Omega \approx \frac{3}{4} \frac{\beta A^2}{\Olin}= -\frac{9}{2} \dOspring \frac{g_0^2}{\kappa^2}A^2 \label{eq:SI:chiNL_shift}
\end{align}
that describes this nonlinear contribution to the spring shift of the \emph{probe} susceptibility, valid under the assumption $\dONLchi \ll \Omega$. This equation is given in the main text as Eq.~\eqref{eq:chiNL_shift}.
Although Eq.~\eqref{eq:SI:chiNL_shift} is derived for a coherent probe tone, it equally predicts the nonlinearly shifted response to weak, incoherent thermal forces, as shown in Fig.~\ref{fig:2}.

\section{Effective nonlinear interactions}
In this section, we present a detailed analysis of the optical backaction acting between a pair of mechanical modes coupled to a common cavity. We start from the two equations of motion 
\begin{align}
    \ddot{z}_j =  - \OlinN{j} ^2 z - \gamma_j \dot{z}_j + f_{\text{opt},j}
\end{align}
for the normalized amplitudes $z_j = x_j / x_{\text{zpf},j}$ of the two resonators $j=1,2$. We write the normalized optical force, arising for now from a single pump laser, as
\begin{align}
    f_{\text{opt},j} = \frac{\hbar G \nc}{m_j x_{\text{zpf},j}} = 2  \OlinN{j} g_{0,j} \nc,
\end{align}
where $\nc$ is total intracavity photon number. 
This is given by
    $\nc = \nmax h(u_0 + \delta u)$,
where, importantly, the normalized optomechanical shift in the cavity frequency 
\begin{align}
    \delta u = \frac{2}{\kappa} \sum_j g_{0,j} z_j
\end{align}
now depends on the displacement $z_j$ of both resonators.

We first consider driving by a single pump laser tuned to either of the maximal spring shift points $u_0 = u_\pm = \pm1/\sqrt{3}$.
Expanding $\nc$ to third order in $\delta u$ then yields
\begin{align}
    \nc &= \nmax \left[ h(u_\pm) + h'(u_\pm) \delta u + \frac{1}{6} h'''(u_\pm) (\delta u)^3 \right] \\
    &= \nmax \left[ h(u_\pm) + h'(u_\pm) \left(\delta u - \frac{3}{4} (\delta u)^3\right) \right],
\end{align}
where we have used $h''(u_\pm) = 0$ and $h'''(u_\pm) = -(9/2)h'(u_\pm)$.
We plug this into the equation of motions to obtain
\begin{align}
    \ddot{z}_j =  &- \OlinN{j} ^2 z - \gamma_j \dot{z}_j + 2\OlinN{j} g_{0,j} \\
    &\times \nmax \left[ h(u_\pm) + h'(u_\pm) \left(\delta u - \frac{3}{4} (\delta u)^3\right) \right]
\end{align}
Again, we remove the static radiation pressure contribution by redefining the displacement origin and get
\begin{align}
    \ddot{z}_j =  &- \OlinN{j} ^2 z - \gamma_j \dot{z}_j + 2\OlinN{j} g_{0,j} \nonumber\\
    &\times \nmax  h'(u_\pm) \left(\delta u - \frac{3}{4} (\delta u)^3\right). \label{eq:SI:eom_z2}
\end{align}

We analyse the optical force term linear in $\delta u$,
\begin{align}
    f_{\text{opt},1}^{(1)} &= 2\OlinN{1} \frac{2 g_{0,1} \nmax h'(u_\pm)}{\kappa} \left(g_{0,1} z_1 + g_{0,2} z_2 \right) \\
    &= - 2\OlinN{1} \dOspring_1 \, z_1 - 2\OlinN{1} \sqrt{\dOspring_1 \dOspring_2} \, z_2, \label{eq:SI:cpl_fopt_lin}
\end{align}
that acts on resonator $1$---the equivalent for resonator $2$ is obtained by exchanging indices $1 \leftrightarrow 2$. The first term in Eq.~\eqref{eq:SI:cpl_fopt_lin} represents the intra-resonator optical spring $k_{\text{opt},1} = 2 m_1 \OlinN{1} \dOspring_1$. Meanwhile, the cross-term in the optical force, expressed in physical units as
\begin{align}
    F_{\text{opt},1}^{(1,\leftrightarrow)} &= m_1 x_{\text{zpf},1} f_{\text{opt},1}^{(1)}
    = -2 m_1 \frac{x_{\text{zpf},1}}{x_{\text{zpf},2}} \OlinN{1} \sqrt{\dOspring_1 \dOspring_2} \, x_2 \nonumber \\
    &= -\sqrt{2 \OlinN{1} m_1 \dOspring_1} \sqrt{2 \OlinN{2} m_2 \dOspring_2} \, x_2 \nonumber \\
    &= -\sqrt{k_{\text{opt},1} k_{\text{opt},2}} \, x_2 = -\kcross \, x_2
\end{align}
represents an inter-resonator optical spring. This spring dynamically couples the displacements $x_1, x_2$ with spring constant $\kcross = \sqrt{k_{\text{opt},1} k_{\text{opt},2}}$.

For a pair of high-quality mechanical resonators separated by a large frequency difference $\Delta\Omega_{12} = \Omega_1 - \Omega_2 \gg \gamma_j$, as we have in our experiment, inducing a static cross-resonator spring $\kcross$ has virtually no effect on their dynamics. We can, however, stimulate frequency conversion between the two modes by modulating $\kcross$ at $\Delta\Omega_{12}$. In experiment, this is achieved by modulating the intensity $\Pin = P_{\text{in},0} (1 + c_\text{h} \cos(\Delta\Omega_{12} t))$ of the spring laser.

After absorbing the intra-resonator spring shifts into the mechanical frequencies $\Oshifted_j = \OlinN{j} + \dOspring_j$, neglecting the cubic term for now and removing terms that are not (mutually) resonant, we arrive at
\begin{align}
    \ddot{z}_1 = -\Oshifted_1^2 z - \gamma_1 \dot{z}_1 - 2\OlinN{1} c_\text{h}\cos(\Delta\Omega_{12} t) \sqrt{\dOspring_1 \dOspring_2} \, z_2
    \label{eq:SI:cpl_mod_lin_z}
\end{align}
For convenience, we write the two displacements as
\begin{align}
    z_j(t) = \Re[ a_j(t) e^{-\rmi \Omega_j t} ] = \frac{a_j(t) e^{-\rmi \Omega_j t} + a_j^*(t) e^{\rmi \Omega_j t}}{2}
\end{align}
in terms of the complex amplitudes $a_j(t)$. We plug these into the equation of motion~\eqref{eq:SI:cpl_mod_lin_z}, neglect the terms that scale with $\gamma_j / \Oshifted_j, J/\Oshifted_j \ll 1$ and remove the counter-rotating terms, to obtain the coupled differential equations
\begin{align}
\begin{aligned}
    \dot{a}_1 &= -\frac{\gamma_1}{2} a_1 - \rmi \frac{c_\text{h} \sqrt{\dOspring_1 \dOspring_2}}{2} a_2 \\
    \dot{a}_2 &= -\frac{\gamma_2}{2} a_2 - \rmi \frac{c_\text{h} \sqrt{\dOspring_1 \dOspring_2}}{2} a_1 \label{eq:SI:cpl_lin_eoms}
\end{aligned}
\end{align}
for the complex amplitudes. From these, we immediately identify the linear coupling rate
\begin{align}
    J = \frac{c_\text{h} \sqrt{\dOspring_1 \dOspring_2}}{2}.
\end{align}

Next, we analyse the nonlinearity in the cross-resonator spring $\kcross$, induced by the cubic term $\propto (\delta u)^3$ in Eq.~\eqref{eq:SI:eom_z2}. Before doing so, we isolate this (cross-Kerr) nonlinearity from the intra-resonator Duffing nonlinearity by driving the cavity with two spring lasers $a$ and $b$ at opposite detunings $u_0^a = u_- = -1/\sqrt{3}$, $u_0^b = u_+ = 1/\sqrt{3}$.
Noting that the difference between $u_-$ and $u_+$ is larger than any mechanical frequency, we can neglect the optical beating between the two lasers and decompose the total photon number 
\begin{align}
    \nc = \nc^a + \nc^b = \nmax^a h\left(u_- + \delta u\right) + \nmax^b h\left(u_+ + \delta u\right) \label{eq:SI:cpl_twolaser_nc1}
\end{align}
into two independent populations $\nc^{a,b}$, one sustained by each laser. 

To induce resonant coupling between the mechanical modes, we modulate one of the laser intensities. Here we modulate the positive spring laser intensity $\nmax^a = \nmax (1 + f(t))$ while keeping the negative spring laser at fixed intensity $\nmax^b = \nmax$, equal to the average $\nmax^a$. After expanding Eq.~\eqref{eq:SI:cpl_twolaser_nc1} to third order, we find that in the total photon number
\begin{align}
    \nc &= \nmax 2h(u_\pm)  \nonumber \\
    &+ \nmax f(t)\left[h(u_+) + h'(u_+) \delta u + \frac{1}{6} h'''(u_+) (\delta u)^3 \right] \nonumber \\
    &= \nmax 2h(u_\pm)  \\
    &+ \nmax f(t)\left[h(u_+) + h'(u_+) \left(\delta u - \frac{3}{4} (\delta u)^3  \right) \right] \nonumber
\end{align}
the static linear and cubic spring shift terms cancel, leaving only the modulated part. 

We plug this into the equation of motion, after discarding the static radiation pressure contribution. Moreover, as the modulation $f(t) = c_\text{h} \cos(\Delta\Omega_{12} t)$ is not resonant with any of the mechanical frequencies $\Omega_1, \Omega_2$, we also discard the modulated direct radiation pressure contribution. We obtain
\begin{align*}
    \ddot{z}_1 =  &- \OlinN{1} ^2 z_1 - \gamma_1 \dot{z}_1 + 2\OlinN{1} g_{0,1} \\
    &\times f(t) \nmax h'(u_\pm) \delta u \left[  1 - \frac{3}{4} (\delta u)^2 \right] \\
    =  &- \OlinN{1} ^2 z_1 - \gamma_1 \dot{z}_1 - 2\OlinN{1} f(t) \frac{2g_{0,1} \nmax h'(u_\pm)}{\kappa} \\
    &\times  \left[ g_{0,1} z_1 + g_{0,2} z_2\right] \left(1 - \frac{3}{4} (\delta u)^2 \right)
\end{align*}
and the equivalent for $z_2$ after exchanging indices $1\leftrightarrow 2$. This can be conveniently expressed in terms of the linear spring shifts as
\begin{align}
    \ddot{z}_1 = &- \OlinN{1} ^2 z_1 - \gamma_1 \dot{z}_1 - 2\OlinN{1} f(t) \sqrt{\dOspring_1} \nonumber \\
    &\times  \Bigg[ \sqrt{\dOspring_1} z_1 \left(1-\frac{3}{\kappa^2}\left(g_{0,1} z_1 + g_{0,2} z_2\right)^2\right) \nonumber \\
    &\quad+ \sqrt{\dOspring_2} z_2 \left(1-\frac{3}{\kappa^2}(g_{0,1} z_1 + g_{0,2} z_2)^2\right) \Bigg] 
\end{align}

Again, we express the two displacements in terms of the complex amplitudes $a_j(t)$, employing the computer algebra software \texttt{Mathematica} to plug these into the coupled nonlinear equation of motion. After neglecting counter-rotating terms and terms scaling with $\gamma_j / \Omega_j, J / \Omega_j \ll 1$, we obtain the dynamical equations
\begin{align}
    \dot{a}_1 = &-\frac{\gamma_1}{2}a_1 \label{eq:SI:cpl_a_evo_nonlin} \\
    &- \rmi J a_2(t) \left[1 - \frac{9}{4}\left(\frac{g_{0,1}^2}{\kappa^2}|a_1(t)|^2 + \frac{g_{0,2}^2}{\kappa^2}|a_2t)|^2\right)\right] \nonumber \\
    &+ \rmi J a_1(t)\frac{9}{2} \left[\frac{g_{0,1}^2}{\kappa^2} \Re[a_1^* a_2]\right] \nonumber \\
    \dot{a}_2 = &-\frac{\gamma_2}{2}a_2 \nonumber \\
    &- \rmi J a_1(t) \left[1 - \frac{9}{4}\left(\frac{g_{0,1}^2}{\kappa^2}|a_1(t)|^2 + \frac{g_{0,2}^2}{\kappa^2}|a_2t)|^2\right)\right] \nonumber \\
    &+ \rmi J a_2(t)\frac{9}{2} \left[\frac{g_{0,2}^2}{\kappa^2} \Re[a_2^* a_1]\right]. \nonumber
\end{align}
for the complex amplitudes. Importantly, in each second term we recognize a coupling rate
\begin{align}
    J_\text{NL} = J \left[1 - \frac{9}{4}\left(\frac{g_{0,1}^2}{\kappa^2}|a_1(t)|^2 + \frac{g_{0,2}^2}{\kappa^2}|a_2t)|^2\right)\right]
\end{align}
that includes a nonlinear correction depending on the squared amplitudes $|a_1(t)|^2$ and $|a_1(t)|^2$.

We initialize the nonlinear ringdown experiment in Fig.~\ref{fig:4} with  excitation in one mode only.
The complex amplitude $a_1(t)$ and $a_2(t)$ then remain $\pi/2$ out of phase throughout, so that the correlation $\Re[a_2^* a_1] = 0$ vanishes. The third term in each equation~\eqref{eq:SI:cpl_a_evo_nonlin} is thus identically zero, leaving us only with a nonlinearly shifted coupling rate $J_\text{NL}$.

\section{Simulating geometric nonlinearity}
\begin{figure}
    \centering
    \includegraphics{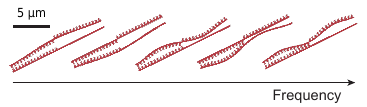}
    \caption{\textbf{Nanobeam flexural modes.} Simulated modeshapes for the first five flexural mechanical modes. All experiments in this work employed the fourth mode, with the mode coupling experiments additionally employing the third mode.}
    \label{fig:si:mode_shapes}
\end{figure}

We estimate the intrinsic geometric nonlinearity from finite element modelling in \texttt{COMSOL Multiphysics}. To do so, we first simulate the linear modeshape of the relevant flexural eigenmode of the device geometry, as shown in Fig.~\ref{fig:si:mode_shapes}. We then perform a series of stationary analyses that include geometric nonlinearity. In this sweep, we impose the linear modeshape scaled by an increasing amplitude factor on the device structure, while for every amplitude, we integrate the total potential energy stored in the strain of the nanobeam~\cite{Hirsch2026tutorial}. This results in a potential energy $U(x) = \frac{1}{2}k_1 x^2 + \frac{1}{4} k_3 x^4$ that is quadratic-quartic in the mode displacement $x$, where $k_1$ is the linear spring constant and $k_3$ the coefficient for the cubic restoring force. 

We fit the potential $U(x)$ obtained from the simulations to extract $k_1$ and $k_3$. We verify that $k_1$, when combined with the effective mass $m_\text{eff}$ obtained from the same simulation, produces the correct eigenmode frequency $\Olin = \sqrt{k_1 / m_\text{eff}}$. We convert $k_3$ into a cubic coefficient using the relation
\begin{align}
    \beta = \frac{k_3}{m_\text{eff}} \xzpf^2
\end{align}
to produce an estimated value of $\beta \approx 300$~Hz$^2$.

\section{Experimental methods}
The experimental set-up, methods, and device employed here have been described in detail in refs.~\cite{delPino2022nonhermitian,Wanjura2022quadrature,Slim2024optomechanical}. We briefly summarize the relevant aspects below.

\subsection{Optomechanical system}
The nano-optomechanical device under study comprised a suspended, sliced silicon nanobeam hosting a photonic crystal cavity and fabricated in-house using the methods described in~\cite{delPino2022nonhermitian}. The optical cavity coupled simultaneously to several non-degenerate flexural modes of the beam, whose spatial profiles are shown in Fig.~\ref{fig:si:mode_shapes}. All experiments in this work employed the fourth mode, with the mode coupling experiments additionally employing the third mode.

Optical access was provided by free-space coupling, achieved by focusing a laser at normal incidence onto the device. Direct reflections were suppressed and cavity-emitted light was collected using a cross-polarized detection scheme. The device was mounted inside a room-temperature vacuum chamber (pressure $2\times10^{-6}$~mbar) and positioned at the laser focus using a piezo-actuated precision stage. All quoted optical and mechanical parameters were characterized following the methods described in~\cite{delPino2022nonhermitian}.

\subsection{Excitation and readout}
Spring shifts were induced by pumping the cavity with a strong `spring' laser (Toptica CTL 1500/1550). Depending on the experiment, a secondary control laser was either used as a `force' laser, or as an additional, opposing `spring' laser. When used as a weak force laser, it was tuned to the cavity resonance with its intensity modulated to provide radiation pressure driving of the mechanical resonator. When used as an additional spring laser, its power was matched to the primary spring laser but operated at the opposite detuning to cancel the (nonlinear) spring effect. In that case, coherent radiation pressure driving was provided by weakly modulating either of the spring lasers instead. In the multimode coupling experiments, the intensity of a spring laser was modulated at the difference frequency of the resonator pair. 

Electronic signals to drive or couple mechanical modes were generated by a high-frequency lock-in amplifier (LIA; Zurich Instruments UHFLI), amplified (Mini-Circuits ZHL-32A+ with 9 dB attenuation) and imprinted on the appropriate laser using fiber coupled electro-optic modulators (Thorlabs LN81S-FC and Covega Mach-10 056)

The instantaneous displacement $x_j(t)$ of both mechanical resonators was transduced onto the reflected intensity of light from a tertiary `detect' laser (Toptica CTL 1550, incident power $P_\text{det} = 1$~mW), far detuned from the cavity resonance ($\Delta_\text{det} = -2.5\kappa$,). 
The reflected detection light was spectrally isolated from control laser light using a tunable band-pass filter (DiCon) before being directed onto a fast, AC-coupled photodiode (New Focus 1811).
To extract individual resonator signals, the photodiode output was demodulated by the LIA with a 3-dB-bandwidth of $50$~kHz (third order filter). The resulting complex amplitudes $a_j(t)$ were normalized to the detector voltage level corresponding to each resonator’s zero-point motion $x_\text{zpf}$, as obtained from thermally driven reference measurements without any modulations. 

For time-resolved ring-down measurements, mechanical driving and coupling signals were continuously output by the LIA on separate output channels, which were then routed through individual radio frequency (RF) switches (Mini-Circuits ZYSWA-2-50DR+). An additional signal generator (Siglent SDG series) controlled the switches and triggered LIA data acquisition.

\section{Extracting instantaneous oscillation frequencies}
The complex resonator amplitude $a(t) = A(t)e^{\rmi \phi(t)}/\xzpf$ reported by the LIA encodes the amplitude $A(t)$ of resonator motion, as well as its phase relative to the LIA's electronic local oscillator (LO). The demodulated phase evolves as
\begin{align}
    \phi(t) = \int_{t_0}^t \left[ \Omega(\tau) - \omega_\text{d} \right]\,\mathrm{d}\tau \label{eq:SI:demod_phase}
\end{align}
where $\Omega(t)$ is the instantaneous frequency of the resonator's motion, $\omega_\text{d}$ the LO frequency and $t_0$ the time the measurement was started. By taking the derivative of Eq.~\eqref{eq:SI:demod_phase}, the instantaneous frequency is extracted $\Omega(t) = \phi'(t) + \omega_\text{d}$. This expression was evaluated for the experimental data to obtain the frequency traces shown in Fig.~\ref{fig:1}f, bottom.

\section{Estimating instantaneous coupling rates}
In this section, we introduce a numerical method to estimate the instantaneous coupling $\Jeff(t)$ from the experimentally obtained complex amplitudes $a_j(t)$ of the coupled resonator pair.

The starting point is the coupled system of differential equations
\begin{align*}
    \dot{a}_1 &= -\frac{\gamma_1}{2} a_1 - \rmi \Jeff(t) a_2, \\
    \dot{a}_2 &= -\frac{\gamma_2}{2} a_2 - \rmi \Jeff(t) a_1
\end{align*}
that describes the evolution of a pair of resonators resonantly coupled by an effective instantaneous rate $\Jeff(t)$ (cf. Eq.~\eqref{eq:SI:cpl_lin_eoms}). 
From these equations of motion, we calculate the `excess change' in each resonator's complex amplitude as
\begin{align}
\begin{aligned}
    \dot{a}_1 + \frac{\gamma_1}{2} a_1 &= - \rmi \Jeff(t) a_2 \\
    \dot{a}_2 + \frac{\gamma_1}{2} a_2 &= - \rmi \Jeff(t) a_1.
\end{aligned}
\end{align}
Here, the left hand sides express the portion of the amplitude change $\dot{a}_j$ that is not attributed to the dissipation $\gamma_1 a_1/2$. From these, an estimate for $\Jeff(t)$ can be immediately obtained through dividing by $a_{2,1}$. However, these amplitudes will periodically tend to zero, giving imprecise results at those times when applied to real data.

To solve this, we square both and take the sum
\begin{align}
	\left| \dot{a}_1 + \frac{\Gamma_1}{2} a_1 \right|^2 + \left| \dot{a}_2 + \frac{\Gamma_1}{2} a_2 \right|^2 &= \Jeff(t)^2 \left( |a_1|^2 + |a_2|^2 \right) 
\end{align}
The total energy $|a_1|^2 + |a_2|^2$ does not vanish periodically as the resonators exchange energy, so that $\Jeff(t)^2$ can be estimated reliably during the entire ringdown.

\end{document}